\documentclass[conference]{IEEEtran}
\IEEEoverridecommandlockouts
\usepackage{cite}
\usepackage{amsmath,amssymb,amsfonts}
\usepackage{algorithmic}
\usepackage{graphicx}
\graphicspath{ {figures/} }
\usepackage{textcomp}

\usepackage{listings}
\usepackage{array}
\usepackage{color}
\usepackage{soul}

\newcolumntype{C}[1]{>{\centering\let\newline\\\arraybackslash\hspace{0pt}}m{#1}}

\def\BibTeX{{\rm B\kern-.05em{\sc i\kern-.025em b}\kern-.08em
    T\kern-.1667em\lower.7ex\hbox{E}\kern-.125emX}}
\begin{document}

\bstctlcite{IEEEexample:BSTcontrol}

\title{Dynamic Thresholding Mechanisms for IR-Based Filtering in Efficient Source Code Plagiarism Detection}

\author{
\IEEEauthorblockN{Oscar Karnalim}
\IEEEauthorblockA{\textit{Faculty of Information Technology} \\
\textit{Maranatha Christian University}\\
Bandung, Indonesia \\
oscar.karnalim@it.maranatha.edu}
\and
\IEEEauthorblockN{Lisan Sulistiani}
\IEEEauthorblockA{\textit{Faculty of Information Technology} \\
\textit{Maranatha Christian University}\\
Bandung, Indonesia \\
lisans1601@gmail.com}
}

\maketitle

\begin{abstract}
To solve time inefficiency issue, only potential pairs are compared in string-matching-based source code plagiarism detection; 
wherein potentiality is defined through a fast-yet-order-insensitive similarity measurement (adapted from Information Retrieval) and only pairs which similarity degrees are higher or equal to a particular threshold is selected.
Defining such threshold is not a trivial task considering the threshold should lead to high efficiency improvement and low effectiveness reduction (if it is unavoidable).
This paper proposes two thresholding mechanisms---namely range-based and pair-count-based mechanism---that dynamically tune the threshold based on the distribution of resulted similarity degrees.
According to our evaluation, both mechanisms are more practical to be used than manual threshold assignment since they are more proportional to efficiency improvement and effectiveness reduction.
\newline
\end{abstract}

\section{INTRODUCTION}
Plagiarizing another student's work and claiming it as theirs for completing course assignment is an illegal behavior for students \cite{Simon2018}.
In programming courses, detecting the existence of such behavior is not trivial considering programming assignments are usually given weekly \cite{Kustanto2009} and source code (i.e., submitted assignment form) is easy to be replicated \& modified \cite{Karnalim2017IAENG}.
As a result, several automated plagiarism detection approaches have been developed \cite{Lancaster2004}.

One of the frequently-used approaches for detecting source code plagiarism is based on string-matching algorithm \cite{Karnalim2017IAENG} (where given source codes are converted to token sequences and treated as strings with each token refers to one character).
Despite its effectiveness, such approach takes a considerable amount of time when applied on academic environment;
given source codes should be compared to each other in combinatoric manner and each comparison typically takes either quadratic or cubic time complexity.

A work in \cite{Burrows2007} addresses aforementioned time inefficiency issue by utilizing Information Retrieval (IR) as an initial filter.
Instead of comparing each possible combination pair with a string-matching algorithm, it only compares several pairs which IR-based similarity degree passes a particular threshold.
In such manner, the processing time will be significantly reduced considering not all pairs are compared with string-matching algorithm and IR-based similarity algorithm usually works in linear time complexity.

To date, the threshold used for IR-based filtering is defined without considering two facts: most IR-based similarity degrees are clustered at certain points and the distribution of those degrees varies per source code plagiarism dataset.
As a result, the tendency to incorrectly assign the threshold will be considerably high.
Such incorrect assignment could either remove numerous potential plagiarism pairs (when assigned threshold is overly-high) or lead to limited time efficiency improvement (when assigned threshold is overly-low).

To mitigate the tendency of incorrect threshold assignment, this paper proposes two dynamic thresholding mechanisms that define the threshold based on the distribution of resulted IR-based similarity degrees.
Range-based thresholding mechanism converts raw threshold (i.e., a manually-defined threshold) to a proportion toward the distribution of resulted IR-based similarity degrees.
Whereas, pair-count-based thresholding mechanism converts the raw threshold to a proportion toward the number of compared pairs;
wherein such proportion will be used to exclude pairs with the lowest similarity degrees.

\section{RELATED WORKS}
Source code plagiarism detection can be classified to three categories: attribute-based, structure-based, and hybrid approach \cite{Al-Khanjari2010,Karnalim2017IAENG}.
Attribute-Based Approach (ABA) relies on shared source code characteristics (e.g., token occurrences \cite{Acampora2015}) to suspect plagiarism. 
Numerous similarity measurements have been used in this approach where most of them are adapted from other domains, such as Fuzzy Logic \cite{Acampora2015} and Information Retrieval \cite{Inoue2012,Ohmann2015,Ganguly2017,Cosma2012}.
Structure-Based Approach (SBA) relies on shared source code structure to suspect plagiarism.
Most of its similarity measurements are based on string-matching algorithms \cite{Prechelt2002,Karnalim2016,Duric2013,Karnalim2017IAENG,Rabbani2017} or graph-matching algorithms \cite{Fu2017, Liu2006}.
Hybrid approach combines ABA and SBA to suspect plagiarism.
Such combination aims to get either better effectiveness \cite{ElBachirMenai2010,Engels2007,Poon2012} or efficiency \cite{Burrows2007}.

Effectiveness-oriented hybrid approach benefits from ABA's and SBA's similarity result characteristics.
For example, a work in \cite{ElBachirMenai2010} shows ABA's and SBA's similarity results at once for suspecting plagiarism.
They argue that slight modification is better handled by ABA while the complex one is handled by SBA.
Other two examples are works proposed in \cite{Engels2007} and \cite{Poon2012}. 
These works treat SBA's similarity degree (resulted from string-matching algorithm) as an attribute for ABA---that utilizes learning algorithm \cite{Engels2007} or clustering algorithm \cite{Poon2012}.

In contrast, efficiency-oriented hybrid approach benefits from ABA's fast processing time to perform efficient comparison using SBA; SBA is commonly slow due to its high time complexity.
A work in \cite{Burrows2007} mitigates the number of SBA-compared source code pairs by performing ABA-based filtering beforehand.
A source code pair is only measured using SBA (that utilizes string-matching algorithm in their case) \textit{iff} its ABA-compared similarity degree (resulted from Information Retrieval measurement) passes a particular threshold.
According to their evaluation, such combination could enhance the efficiency of source code plagiarism detection with no extreme reduction on effectiveness. 

Considering aforementioned filtering (which will be referred as IR-based filtering at the rest of this paper) is conducted prior to string-matching algorithm\cite{Burrows2007}, defined threshold for such filtering is crucial to determine the effectiveness and efficiency.
It could lead to ineffectiveness when defined threshold is higher than most pairs' IR-based similarity degrees;
no pairs will be passed to string-matching algorithm (which is responsible to determine plagiarism cases).
On another extreme point, it could also lead to time inefficiency since most pairs will be compared using string-matching algorithm (which takes a considerable amount of processing time) in addition to IR-based similarity measurement.

\section{METHODOLOGY}
This paper proposes two mechanisms that mitigate the tendency of incorrectly assigning threshold for IR-based filtering.
Those mechanisms---which are referred as range-based and pair-count-based mechanism---dynamically tune raw threshold based on the distribution of resulted IR-based similarity degrees.
They are expected to be more practical to be used than manual threshold assignment;
they are more proportional to efficiency improvement and effectiveness reduction. It is important to note that effectiveness reduction is unavoidable in most cases since IR-based measurements are typically less stricter than the string-matching one.

Range-based thresholding mechanism considers raw threshold (i.e., the manually-defined one) as a proportion toward the distribution of resulted IR-based similarity degrees.
It is calculated using (\ref{eq:RM}) where \textit{in} refers to raw threshold and \textit{$sim_{max}$} \& \textit{$sim_{min}$} refer to maximum \& minimum IR-based similarity degrees on given dataset respectively.
For instance, if 50\% is passed as the raw threshold toward a plagiarism dataset which maximum \& minimum IR-based similarity degree are 80\% \& 30\% respectively, range-based thresholding mechanism will exclude all pairs which IR-based similarity degree is lower than 55\%---that is resulted from 30\% + (50\% * (80\% - 30\%)).
\begin{equation}
RM(in) = sim_{min} + (in * (sim_{max} - sim_{min}))
\label{eq:RM}
\end{equation}

In contrast, pair-count-based thresholding mechanism considers raw threshold as a proportion toward the number of compared pairs.
Such proportion will then be used to exclude pairs with the lowest IR-based similarity degrees.
The number of excluded pairs will be calculated as in (\ref{eq:PCM}) where \textit{total\_pairs} refers to the number of compared pairs.
If raw threshold is 50\% and the number of compared pairs is 30, pair-count-based thresholding mechanism will exclude 15 pairs (50\% of 30) with the lowest similarity degrees.
\begin{equation}
PCM(in) = in * total\_pairs
\label{eq:PCM}
\end{equation}

\begin{figure}[htbp]
\centerline{\includegraphics[width=0.25\textwidth]{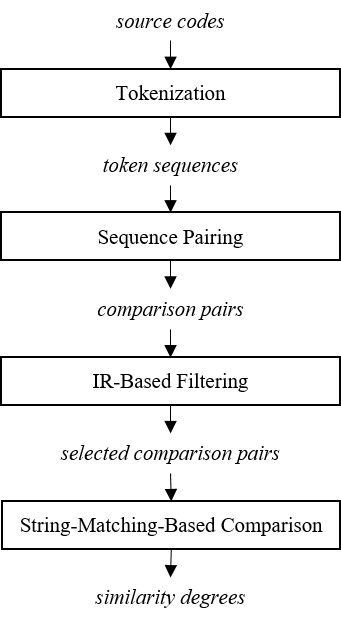}}
\caption{
Efficiency-oriented hybrid source code plagiarism detection used in our study. It contains four consecutive phases.
}
\label{fig:methodology}
\end{figure}

For our case study, both thresholding mechanisms will be applied on an efficiency-oriented hybrid source code plagiarism detection, which works in fourfold (see Fig. \ref{fig:methodology}).
At first, given source codes (where each code represents a student's work) are converted to token sequences using ANTLR \cite{Parr2013} with comment tokens excluded.
Second, comparison pairs are formed by pairing each token sequence with other sequences in combinatoric manner.
For example, if there are three sequences named \textit{A}; \textit{B}; and \textit{C}, their comparison pairs will be \textit{(A,B)}, \textit{(A,C)}, and \textit{(B,C)}.
Third, for each comparison pair, IR-based similarity degree is measured using vector space model and cosine similarity \cite{Croft2010}.
If such degree passes defined threshold (generated by either range-based or pair-count-based mechanism), that pair will be passed to the $4^{th}$ phase.
Otherwise, the pair will be excluded from consideration.
Fourth, string-matching-based similarity degree of passed comparison pairs will be measured using either Running-Karp-Rabin Greedy-String-Tiling (RKRGST) \cite{Wise1995} or Local Alignment (LA) \cite{SmithTempleF;Waterman1981}; where both algorithms have been modified to handle source code tokens instead of characters.
The results of this phase will be used as a guideline to suspect source code plagiarism.

\section{EVALUATION}
Three evaluation metrics will be considered in this study: the number of excluded pairs, reduced asymptotic number of processes (i.e., predicted reduced processing time), and the dissonance degree of excluded pairs (i.e., tendency to exclude potentially-suspected comparison pairs).
The former two are related to time efficiency while the last one is related to effectiveness.

Each evaluation metric will compare three thresholding mechanisms: Range-based thresholding Mechanism (RM), Pair-Count-based thresholding Mechanism (PCM), and Static thresholding Mechanism (SM).
RM and PCM are our proposed mechanisms.
SM, on the other hand, is a manual thresholding mechanism that has been used in \cite{Burrows2007}.
It directly uses raw threshold as similarity degree threshold in IR-based filtering.
For comparison purpose, SM will utilize the same efficiency-oriented hybrid source code plagiarism detection as in RM and PCM.
All thresholding mechanisms will be converted to 11 evaluation scenarios each, varying in terms of initial threshold (starting from 0\% to 100\% with 10\% increase between scenarios).

Evaluation dataset is taken from \cite{Karnalim2018Icoict} by clustering all plagiarized codes per plagiarist, resulting 9 sub-datasets.
Such dataset was initially formed by asking 9 lecturer assistants to plagiarize 7 Java source codes (covering Introductory Programming topics) using 6 plagiarism attack categories (defined in \cite{FaidhiJ.A.W;Robinson1987} with comment \& whitespace modification as the lowest level and logic change as the highest one).
The statistics of all sub-datasets can be seen in Table \ref{tab:subdatasets_statistics}. 
For each sub-dataset, source code pairs are generated by comparing the codes to each other.

\begin{table}[htbp]
\caption{The Statistics of Evaluation Sub-Datasets}
\label{tab:subdatasets_statistics}
\centering
\begin{tabular}{| C{0.02\textwidth}| C{0.08\textwidth}| C{0.08\textwidth}| C{0.08\textwidth}| C{0.08\textwidth}|}
\hline
\bfseries ID & \bfseries Min Token Size & \bfseries Max Token Size & \bfseries Avg Token Size & \bfseries Number of Pairs \\
\hline
P1 & 251 & 1268 & 638.35 & 861 \\ \hline
P2 & 273 & 1499 & 683.92 & 741 \\ \hline
P3 & 269 & 1130 & 614.78 & 820 \\ \hline
P4 & 256 & 1321 & 671.21 & 496 \\ \hline
P5 & 271 & 1240 & 651.62 & 780 \\ \hline
P6 & 369 & 1245 & 689.31 & 861 \\ \hline
P7 & 303 & 1183 & 646.30 & 780 \\ \hline
P8 & 288 & 1195 & 660.91 & 528 \\ \hline
P9 & 323 & 1234 & 644.20 & 780 \\ \hline
\end{tabular}
\end{table}

\subsection{The Number of Excluded Pairs}
The Number of Excluded Pairs (NEP) is defined by counting how many comparison pairs are out of consideration since their IR-based similarity degrees are lower than defined threshold.
It is related to time efficiency since processing time is affected by the number of comparison pairs passed to string-matching-based similarity measurement.

\begin{figure}[htbp]
\centerline{\includegraphics[width=0.5\textwidth]{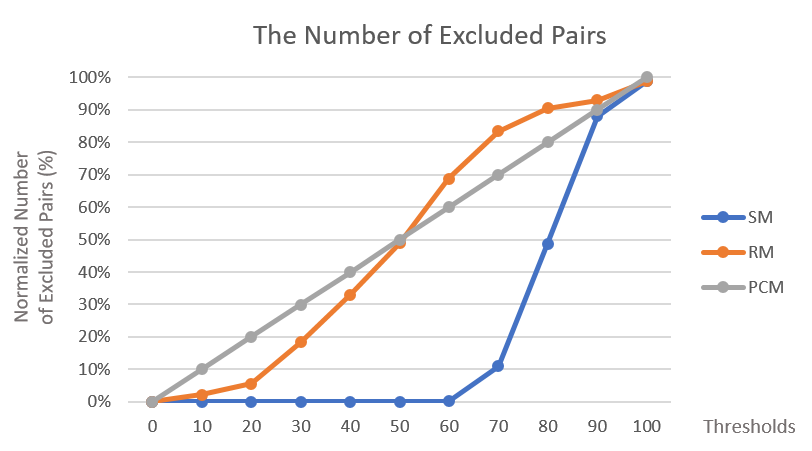}}
\caption{
The number of excluded pairs that is averaged among 9 sub-datasets wherein such number for each sub-dataset is normalized toward total pairs. 
}
\label{fig:excluded_pairs}
\end{figure}

\begin{figure}[htbp]
\centerline{\includegraphics[width=0.5\textwidth]{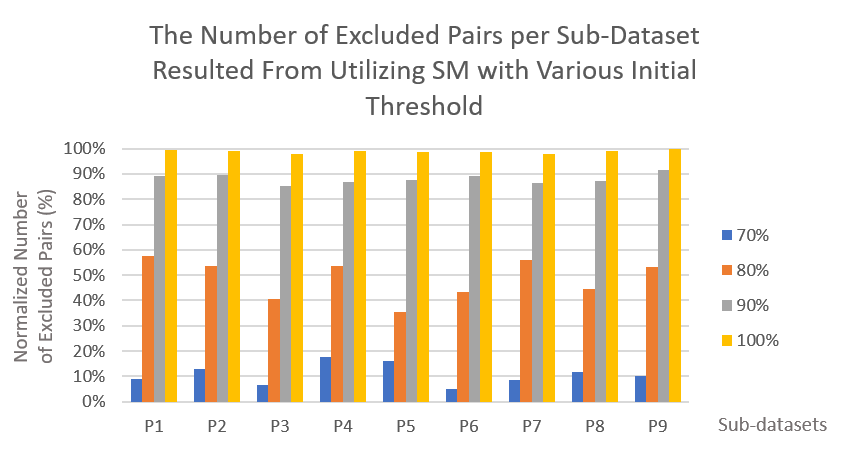}}
\caption{
The number of excluded pairs per sub-dataset from 70\% to 100\% threshold; each number is normalized toward total pairs. 
}
\label{fig:excluded_pairs_sub}
\end{figure}

Fig. \ref{fig:excluded_pairs} shows that SM's scenarios are not proportional to NEP.
It only excludes comparison pairs on limited thresholds (which are 70\% to 100\% in our case).
Such finding is natural considering IR-based similarity degrees are not equally distributed from 0\% to 100\%.
In most occasions, those degrees are clustered at the end of similarity range (more than 50\%).
Further, even in similar scenario, SM excludes various number of comparison pairs among sub-datasets.
As depicted on Fig. \ref{fig:excluded_pairs_sub}, it only generates considerably similar result on 100\% threshold where most pairs are excluded.

PCM's and RM's scenarios, on the other hand, are more proportional to NEP.
Since both PCM and NEP utilize similar metric unit (i.e., the number of pairs), they are completely proportional to each other. 
As seen in Fig. \ref{fig:excluded_pairs}, PCM forms a straight-diagonal graph line as NEP is improved gradually.
RM, in contrast, does not form a straight-diagonal graph line as PCM;
they exclude more pairs at middle thresholds (which are 20\% to 70\%) than the extreme ones since minimum and maximum similarity degrees (used for defining range-based threshold) are usually outliers, that are located far higher or lower than most pairs' similarity degrees.

\subsection{Reduced Asymptotic Number of Processes}
Reduced Asymptotic Number of Processes (RANP) is calculated by subtracting Asymptotic Number of Processes (ANP) prior and upon the implementation of IR-based filtering.
It is related to time efficiency considering ANP is adapted from asymptotic time complexity, a measurement unit for predicting processing time.

ANP for each scenario (per thresholding mechanism) is generated based on (\ref{eq:ANP}) where \textit{P} is a list of comparison pairs and \textit{t} is a defined threshold for IR-based filtering.
It sums up ANP for all comparison pairs where each of them consists ANP for IR-based filtering---see (\ref{eq:IR})---and string-matching-based similarity measurement---see (\ref{eq:SM}).
The former one is the total token size of a comparison pair (where \textit{$p_a$} and \textit{$p_b$} are the token sequences) with an assumption that IR-based filtering takes at most linear complexity.
Whereas, the latter one takes cubic time complexity of the largest size between \textit{$p_a$} and \textit{$p_b$}.
It is based on RKRGST's time complexity since such algorithm is popular to be used in source code plagiarism detection.
It is important to note that $T_{S}$ is only calculated if resulted IR-based similarity degree is higher or equal to defined threshold.
\begin{equation}
ANP(P,t) = \sum_{i=1}^{n} (T_{IR}(P_i) + T_{S}(P_i, t))
\label{eq:ANP}
\end{equation}
\begin{equation}
T_{IR}(p) = |p_a| + |p_b|
\label{eq:IR}
\end{equation}
\begin{equation}
T_{S}(p,t) = 
\begin{array}{ll}
(max(|p_a|,|p_b|))^3 & if\ sim_{IR}(p) >= t \\
\ 0 & otherwise\\
\end{array}
\label{eq:SM}
\end{equation}

\begin{figure}[htbp]
\centerline{\includegraphics[width=0.5\textwidth]{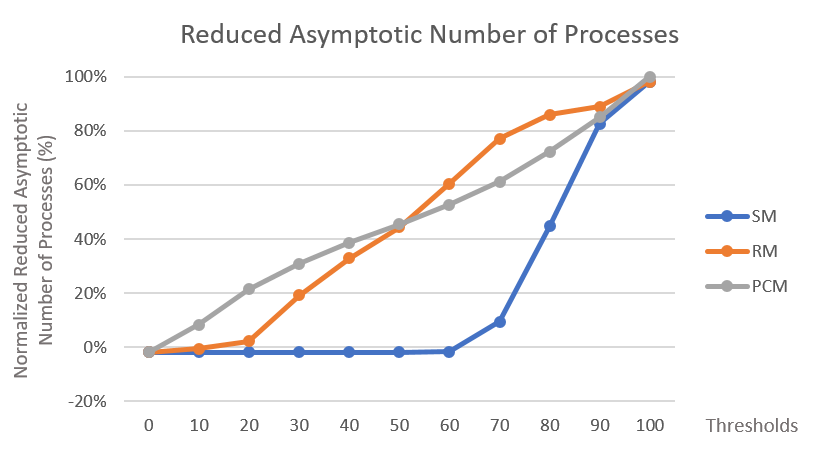}}
\caption{
Reduced asymptotic number of processes. Value for each threshold is averaged among 9 sub-datasets.
}
\label{fig:RANP}
\end{figure}

IR-based filtering is applied prior string-matching-based comparison measurement. 
Hence, it will lead to more processes than conventional approach (i.e., a detection  with string-matching measurement only) when few (or no) comparison pairs are excluded.
As depicted in Fig. \ref{fig:RANP}, some scenarios yield RANP below than 0\% since they exclude few comparison pairs.

RANP result (see Fig. \ref{fig:RANP}) shares similar pattern as NEP result (see Fig. \ref{fig:excluded_pairs}): 
PCM is the most proportional one toward given metric, followed by RM and SM.
Such resemblance occurs due to the fact that both metrics (i.e., RANP and NEP) are related to each other: more excluded pairs leads to fewer involved processes.
However, their results are not exactly similar since token size for each involved token sequence (used for calculating ANP) varies.

\subsection{The Dissonance Degree of Excluded Pairs}
Considering SBA is probably more effective than ABA in most occasions \cite{Verco1996,Karnalim2018icast}, it is important not to exclude comparison pairs with high SBA-measured similarity degrees (i.e., high string-matching-measured similarity degrees) through ABA-based filtering (i.e., IR-based filtering).
The Dissonance degree of Excluded Pairs (DEP) measures how many comparison pairs with high string-matching-based similarity degrees are excluded as a result of IR-based filtering.
Higher DEP refers to lower effectiveness since it means numerous pairs with high string-matching-based similarity degrees are excluded.

DEP is calculated in threefold.
First of all, comparison pairs are ranked in descending order toward two perspectives: IR-measured and string-matching-measured similarity degrees.
Second, per perspective, comparison pairs are assigned with a rank each where the $1^{st}$ rank is assigned to a pair with the highest similarity degree.
Third, DEP is defined based on (\ref{eq:DEP}) where \textit{E} refers to \textit{n} excluded pairs and \textit{$Rank_{IR}$} \& \textit{$Rank_{SIM}$} are the ranks of given pair from IR-measured \& string-matching-measured similarity degree perspective.
Using such equation, the highest and lowest DEPs occurs when half comparison pairs are excluded.
The highest one occurs when all of them are top-half string-matching-measured ranks while the lowest one occurs when all of them are the bottom-half ranks.
For comparison purpose, resulted DEP will be normalized based on its possible lowest score. 
\begin{equation}
DEP(E) = \sum_{i=1}^{n} Rank_{IR}(e_i) - \sum_{i=1}^{n} Rank_{SIM}(e_i) 
\label{eq:DEP}
\end{equation}

\begin{figure}[htbp]
\centerline{\includegraphics[width=0.5\textwidth]{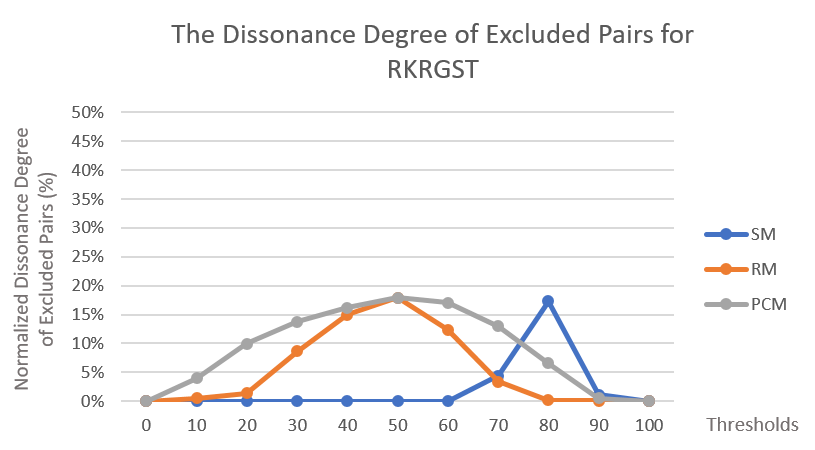}}
\caption{
The dissonance degree of excluded pairs for RKRGST. Value for each threshold is averaged among 9 sub-datasets.
}
\label{fig:dis_rkrgst}
\end{figure}

When RKRGST is used in string-matching-based similarity measurement, SM only yields a non-zero DEP when given threshold is higher or equal to 70\% (see Fig. \ref{fig:dis_rkrgst}).
In contrast, RM and PCM yield DEP in more gradual manner.
Their DEP is increased progressively as given threshold gets closer to 50\%.

When compared to each other, RM is more effective than PCM; 
it generates lower DEP on most scenarios (see Fig. \ref{fig:dis_rkrgst} where RM's graph line is drawn below PCM's).
In average, RM generates 3.58\% less DEP.

All thresholding mechanisms are considerably effective for excluding irrelevant pairs when RKRGST is used in string-matching-based similarity measurement.
Their highest DEP (at 50\% threshold) is extremely low when compared to DEP for worst case scenario; 
it only takes about 17\% of worst case scenario's.
Further observation shows that most excluded pairs have considerably similar rank when perceived from IR-measured and string-matching-measured similarity degree perspectives;
similarity degrees resulted from IR-based similarity measurement (i.e., cosine similarity) strongly correlates with RKRGST's.
According to Pearson correlation, both of them share 0.79 of 1 correlation degree.

\begin{figure}[htbp]
\centerline{\includegraphics[width=0.5\textwidth]{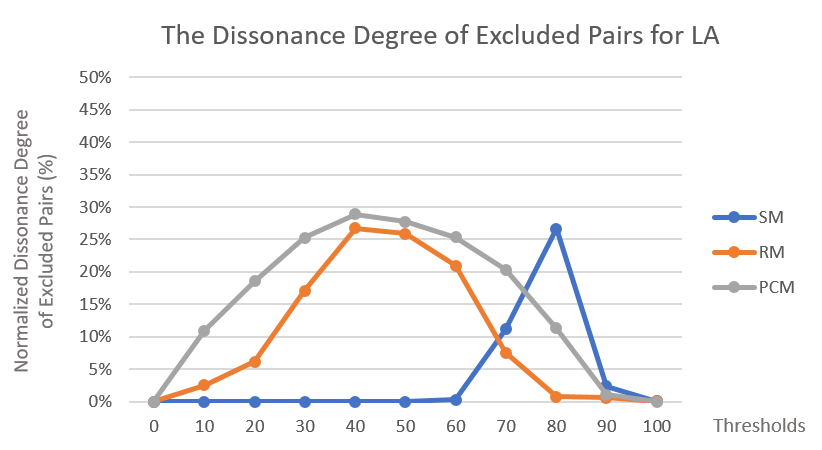}}
\caption{
The dissonance degree of excluded pairs for LA. Value for each threshold is averaged among 9 sub-datasets.
}
\label{fig:dis_la}
\end{figure}

Fig. \ref{fig:dis_la} shows that DEP resulted from using LA as string-matching-based similarity measurement shares similar pattern as using RKRGST.
Such finding is natural since, when measured using Pearson correlation, LA's similarity degrees strongly correlate with RKRGST's (which is 0.78 of 1 correlation degree).
Despite high resemblance, LA favors RM than PCM more.
They generate 5.55\% DEP difference in LA while only 3.58\% difference resulted in RKRGST.
Another finding is that, in general, LA is less effective than RKRGST when used as a part of efficiency-oriented hybrid source code plagiarism detection. 
It generates 1.62\%, 4.45\%, and 6.42\% more DEP on SM, RM, and PCM respectively.

\section{CONCLUSION AND FUTURE WORK}
In this paper, two dynamic thresholding mechanisms for IR-based filtering (i.e., range-based and pair-count-based thresholding mechanism) are proposed.
According to our evaluation, such mechanisms are more practical to be used since they are more proportional than static thresholding mechanism in terms of excluding comparison pairs, reducing number of processes, and mitigating dissonance degree (i.e., a tendency to remove pairs with high string-matching-measured similarity degrees).
When compared to each other, pair-count-based thresholding mechanism is more proportional to the number of excluded pairs and reduced asymptotic number of processes.
However, range-based thresholding mechanism is more effective since it generates less dissonance degree.
The implementation of range-based thresholding mechanism has been implemented in our other work (which details can be seen in \cite{Sulistiani2018}).

For future works, we plan to evaluate whether the effectiveness and efficiency of proposed thresholding mechanisms are consistent among various algorithms for IR-based filtering.
Further, we also plan to check whether frequently-used features on source code plagiarism detection (such as method linearization \cite{Karnalim2017ICSESS,Karnalim2018}) enhance the effectiveness and efficiency of proposed thresholding mechanisms. 


\IEEEtriggeratref{25}
\bibliographystyle{IEEEtran}
\bibliography{IEEEabrv,references}

\end{document}